\documentclass[aps,prl,twocolumn,superscriptaddress,showpacs,english]{revtex4-1}

\usepackage[T1]{fontenc}
\usepackage{babel}
\usepackage{amsmath}
\usepackage{amssymb}
\usepackage{wasysym}
\usepackage{graphicx}
\usepackage{xcolor}
\usepackage{braket}
\usepackage{multirow}

\usepackage[linktocpage=true,
  colorlinks=true, 
  pdfborder={0 0 0},
  linkcolor=blue,
  citecolor=red,
  filecolor=yellow,
  urlcolor=blue,
  bookmarks,
  pdfauthor={},
]{hyperref}

\newcommand{\Rome}{Dipartimento di Fisica, Sapienza Universit\`a di Roma, 00185 Roma, Italy}

\newcommand{\sh}{SH$_3$}

\newcommand{\mg}{MgB$_2$}
\newcommand{\lah}{LaH$_{10}$}

\newcommand{\tc}{$T_\text{c}$}

\newcommand{\ep}{\textit{e-ph}~}

\definecolor{mygreen}{rgb}{0.0, 0.6, 0.0}

\begin{document}

\title{Viewpoint: the road to room-temperature conventional superconductivity}

\author{Lilia Boeri} \email{lilia.boeri@uniroma1.it}
\affiliation{\Rome} 
\author{Giovanni B. Bachelet} \email{giovanni.bachelet@uniroma1.it}            
\affiliation{\Rome}

\date{\today}

\begin{abstract}
It is a honor to write a contribution on this memorial for Sandro Massidda. For both of us, at different stages of our life, Sandro was first and foremost a friend.
We both admired his humble, playful and profound approach to life and physics.
In this contribution we describe the route which permitted to meet
a long-standing challenge in solid state physics, i.e. room temperature superconductivity. In less than 20 years
the \tc\; of {\em conventional} superconductors, which in the
last century had been widely believed to be limited to 25 K,
was raised from 40 K in \mg\; to 265 K in \lah.
This discovery was enabled by the development and application of computational methods
for superconductors, a field in which Sandro Massidda played a major role.
\end{abstract}

\pacs{~}
\maketitle

Since K.H. Onnes discovered in 1911 that a sample of mercury, cooled below a critical temperature (\tc) of 4~K,
exhibits a vanishing resistivity, it became immediately clear that such a {\em superconductivity}, realized at ambient conditions,
could have spectacular 
electrical network applications.  A related unique property of superconductors, perfect diamagnetism, is equally attractive, because quantum
levitation paves the road to futuristic scenarios, such
as levitating vehicles.

However, more than 100 years after the original discovery, none of these large-scale applications has advanced to a point where it is economically viable. 
Currently, the only applications of superconductors are found in devices and
facilities whose cost is not an issue: superconducting magnets are used in 
large-scale particle accelerators and storage rings, diagnostical devices, antennas etc. 
The most severe obstacle to cost-sensitive applications are
the prohibitive refrigeration costs required to cool the existing
materials below their critical temperatures.
The only compounds which superconduct above the liquid nitrogen (N$_2$)
boiling point,
the high-\tc\; cuprates, have several characteristics which make them not
suitable for large scale applications: being brittle, they have high
manufacturing costs; the presence of magnetism in the phase diagram
and the  {\em exotic} symmetry of the superconducting
gap make them sensitive to grain boundaries.~\cite{Gurevich_NATMAT_2011}

As a result, within today's technological applications, the most used superconductors are simple intermetallic alloys, with \tc's in the range
of 10-30 K. In addition to being much cheaper and easier to manipulate, these
superconductors are also easier to describe theoretically: unlike the {\em exotic} cuprates,
they  are described to a high degree of accuracy by the theory of conventional phonon-mediated superconductivity, developed in the 60's; the progress of current {\em ab-initio} methods to calculate the electron-phonon coupling in the last 20 years is such that the normal- and superconducting-state properties  of actual materials can now be computed to a high degree of accuracy based on the sole knowledge of their chemical composition and crystal structure; actually, once the formula unit is given, the equilibrium crystal structure
itself may be predicted, to a great degree of accuracy, using a combination of {\em ab-initio} structural energies and modern optimization methods, a valuable theoretical-computational tool which becomes almost indispensable when one or some of the chemical components are light elements which are hard to locate by X-ray diffraction.

The availability of quantitative methods to describe the electronic, vibrational, and structural properties
of materials has granted not only an accurate understanding of many existing superconductors,
but also the exploration and prediction of hypothetical, new ones, up to providing, last year, the solution to a
puzzle previously deemed impossible: a room-temperature superconductor.
In the summer of 2018, two different groups reported the
discovery of a \tc\; of 260~K in a sample of lanthanum hydride (\lah),
pressurized to Megabar pressures
($\sim 150$ GPa) ~\cite{La_EXP1,La_EXP2,La_EXP3}, and
already three years before another high-pressure hydride,
\sh, had set the all-time superconducting record with a \tc\; of
203 K~\cite{H:DrozdovEremets_Nature2015}; in both cases
their stability and superconductive nature, and to a very good approximation their  \tc ,
had been predicted by {\em ab-initio} calculations.~\cite{Duan_SciRep2014,MA_REHX_PRL_2017}

The discovery of high-\tc\; superconductivity in high-pressure hydrides
does not solve any technological problem,
since, in daily life, to achieve pressures million times larger than ambient pressure
is much more difficult than refrigerating samples down to a few
K. Nonetheless, the discovery of room-temperature
superconductivity in \lah\; has smashed several
 psychological barriers, showing: ($i$) that superconductivity may be achieved at ambient temperature; ($ii$) that high-\tc\; superconductivity
may be achieved by the conventional (electron-phonon) mechanism;
($iii$) that, using first-principles calculations, new superconductors may
be reliably {\em predicted} (and designed) before experiments.

\begin{figure*}[h!t]
	\includegraphics[width=2.00\columnwidth]{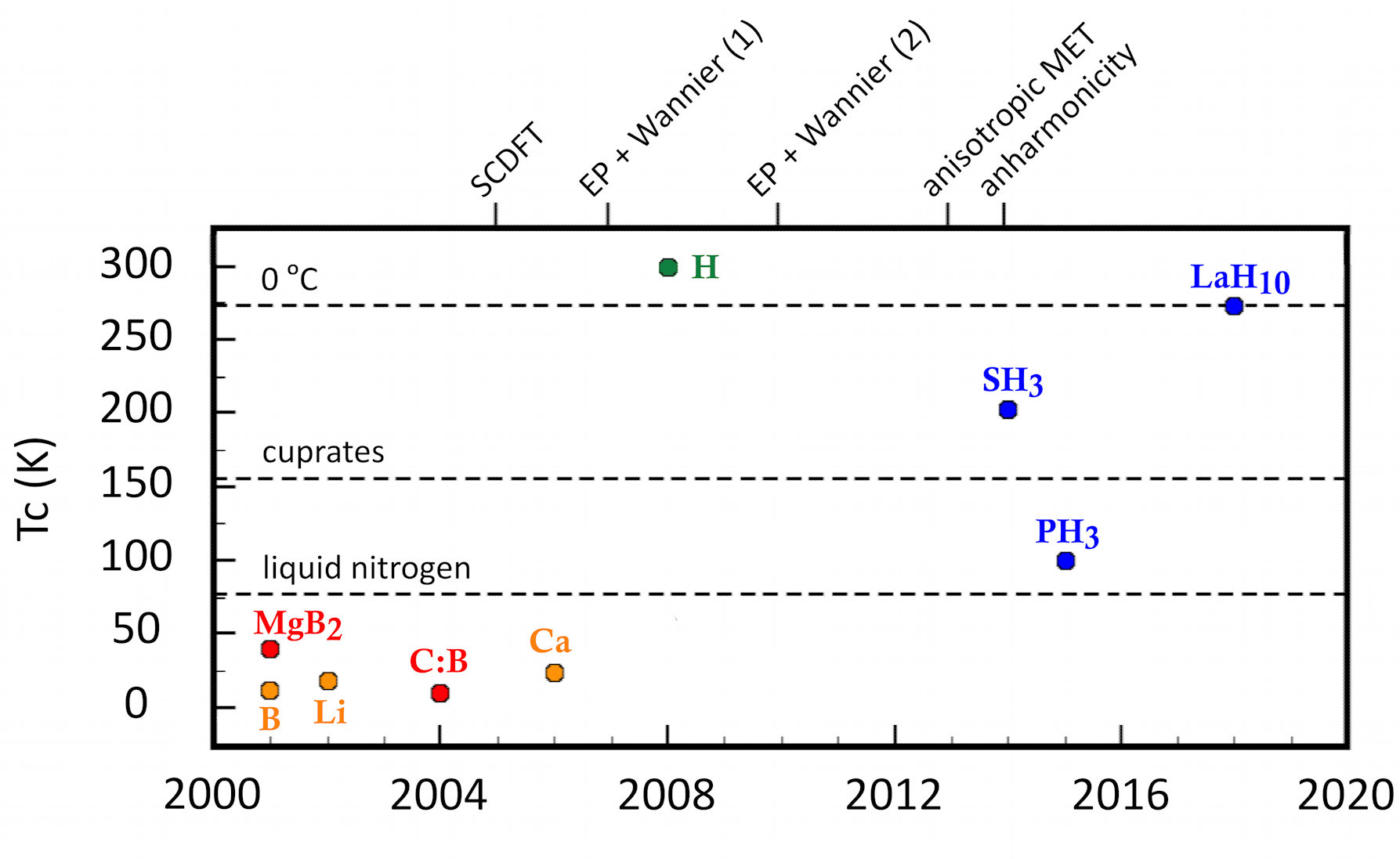}\\
	\caption{The main four milestones on the route to room-temperature
          superconductivity in the 21${st}$ century: discovery of \mg\;
          and other covalent superconductors (red); elemental superconductors at high pressures (orange); theoretical prediction of the phase diagram and
          superconductivity in metallic hydrogen (green); superconductivity in metallic hydrides at high pressure (blue).
          The tics on the top axis mark crucial developments in the field of computational superconductivity, a field pioneered by Sandro Massidda and collaborators:
          Superconducting Density Functional Theory (SCDFT)~\cite{Th:Luders_PRB_2005,Th:Marques_PRB_2005}; \ep interaction with Wannier functions~\cite{DFT:Giustino_PRB_2007,DFT:profeta_PRB_2010}; anisotropic Migdal-Eliashberg Theory~\cite{DFT:giustino_RMP_2017}; self-consistent calculation of anharmonic effects on
phonon spectra~\cite{DFT:Errea_PRB_2014}.
	\label{fig:fig1}}
\end{figure*}

Obviously, the achievement of room-temperature superconductivity
was not a matter of sheer luck, but rather the
result of a long process, which experienced a strong acceleration at the beginning of this
century.
The aim of this viewpoint is to illustrate the last steps
of this process, which took place in the last 20 years.
Out of the rich literature on the subject, we have identified
four milestones, illustrated in Fig.~\ref{fig:fig1}, which marked the route to
the \lah\; discovery. 
\begin{enumerate}
\item The discovery of \mg\; and other covalent superconductors (2001-), red circles.
\item The study of elemental superconductors at high pressures (2002-), orange circles.
\item The theoretical calculation of the phase diagram and \tc{'}s of metallic Hydrogen. (2007-), green circle.
\item The prediction of superconductivity in high-pressure hydrides (2008-), blue circles.
\end{enumerate}
The two last milestones are directly related to
two brilliant intuitions on solid hydrogen and hydrogen-rich hydrides due
to Neil Ashcroft~\cite{H:Ashcroft_PRL1968, H:Ashcroft_PRL2004} which we will discuss later.
It is also interesting to note that, at the start of the route, the experimental
discovery of the remarkable superconducting {\tc} of  {\mg} preceded its 
 understanding in terms of {\em ab initio} electronic bands and phonons, 
whereas, towards the end of the route, theoretical predictions based on 
large-scale {\em ab initio} calculations led experimental groups
to synthesize previously unknown hydrides like {\sh} or {\lah}, eventually
confirming both their stability and their exceptionally high {\tc}.
In other words, the development and extensive application of {\em ab-initio} methods
was so impetuous as to turn them, in less than 20 years, from valuable 
instruments for the interpretation of existing superconductors into reliable
forecasting tools for new superconductors.
Before entering the route leading to the discovery of \lah, we need, however, to rapidly recall
the basic theoretical background of superconductivity, a field to which Sandro, with his students and collaborators, gave an essential contribution.

The simplest microscopic theory of superconductivity, the Bardeen-Cooper-Schrieffer (BCS) theory,
describes the superconducting state as a state of macroscopic quantum coherence,
where electrons form pairs of opposite spin and momentum ({\em Cooper Pairs}) held together by an attractive {\em glue}.
The superconducting state is characterized by the presence of a gap $\Delta$
in the superconducting spectrum, which closes at the critical temperature \tc;
above \tc, conventional superconductors behave as normal metals.
In the original BCS paper, the glue is provided by lattice vibrations (phonons),
but other mediators are in principle possible, such as plasmons, spin fluctuations, charge fluctuations, etc.

The strong-coupling  extension of BCS theory, known as
Migdal-Eliashberg (ME) theory, is a diagrammatic theory for
interacting electrons and bosons~\cite{AllenMitrovic1983}.
At the heart of the theory is the so-called 
 (isotropic)
electron-phonon spectral function, defined as:
\begin{equation}
 \alpha^2 F(\omega) = \frac{1}{N(E_F)} \sum \limits_{\mathbf{k} \mathbf{q},\nu} 
|g_{\mathbf{k},\mathbf{k}+\mathbf{q},\nu}|^2 \delta(\epsilon_\mathbf{k}) 
\delta(\epsilon_{\mathbf{k}+\mathbf{q}}) \delta(\omega-\omega_{\mathbf{q},\nu})~
, 
\label{eq:alpha}
\end{equation}
where $N(E_F)$ is the
DOS at the Fermi level, $\omega_{\mathbf{q},\nu}$ is the phonon frequency of
mode $\nu$ and wave vector $\mathbf{q}$,
and $|g_{\mathbf{k},\mathbf{k}+\mathbf{q},\nu}|$
is the \ep matrix element between two electronic states of wave
vectors $\mathbf{k}$ and $\mathbf{k+q}$ at the Fermi level. 
The expression of the \ep spectral function $\alpha^2F(\omega)$
clearly evidences why superconductors are so hard to predict (or why superconducting properties are so sentitive to small details of the electronic structure
of a given material): 
since in Eq.~(\ref{eq:alpha}) 
the phonon spectrum is weigthed by
the \ep matrix elements  $|g_{\mathbf{k},\mathbf{k}+\mathbf{q},\nu}|$, 
which can be very different for different phonon modes, and since the double delta function
$\delta (\varepsilon_{\mathbf{k}}^n)\delta (\varepsilon_{\mathbf{k+q}}^m)$ restricts the sum of $e\!-\!ph$ matrix elements to electronic states at the Fermi level,
only a small fraction of the full electronic
and phononic spectrum contributes to superconductivity.

To understand and predict new conventional superconductors essentially amounts to explaining or conceiving materials where electrons and phonons are strongly coupled (large $|g_{\mathbf{k},\mathbf{k}+\mathbf{q},\nu}|$),
which can be done with considerable accuracy since all quantities entering the \ep spectral function can be computed  from first-principles within Density Functional Perturbation Theory~\cite{Baroni_RMP2001}.
Once the \ep spectral function is known, the critical temperature of a material can be computed to different degrees of sophistication. The simplest
approach adopts an approximate formula for \tc; a  popular
choice for phonon-mediated superconductors is the 
Mc-Millan-Allen-Dynes formula~\cite{Th:AllenDynes_PRB_1975}
\begin{equation}
\label{eq:McMillan}
  T_c=\frac{\omega_{\log}}{1.2 k_B}\exp\left[-\frac{1.04(1+\lambda)}{\lambda-\mu^{*}(1+0.62\lambda)}\right]\,,
  \,\,\text{where}
\end{equation}
$\lambda\!=\!2 \int d\omega \frac{\alpha^2 F(\omega)}{\omega}$ and $\omega_{\log}\!=\!\exp\left[\frac{2}{\lambda}\int \frac{d\omega}{\omega} \alpha^2 F(\omega)\ln(\omega) \right]$ are
the \ep coupling constant and logarithmic-averaged phonon frequency; 
 $\mu^*$ is the so-called Morel-Anderson pseudopotential,  obtained by
screening the full Coulomb potential up to a characteristic cut-off energy.~\cite{Th:Morel_Anderson_1962}

This approach neglects
effects which may influence the behavior of a superconductor,
such as possible anisotropies of the superconducting gap, strong-coupling
corrections to the \tc\; expression, self-energy effects on the
electronic and phonon spectra, etc.
These effects may be handled from first principles in two ways: ($i$) on one hand,
 the so-called anisotropic Migdal-Eliashberg Theory (AMET), which solves the anisotropic Migdal-Eliashberg equations,
computing the full electronic and phonon self-energies; although very accurate, it cannot be considered
a fully {\em ab-initio} theory, but rather an advanced combination of many-body techniques and first-principles ingredients;  ($ii$)
on the other hand,  the superconducting Density Functional Theory (SCDFT) proposed by
Oliveira, Gross and Kohn in 1988~\cite{Th:Oliveira_PRL_1988} but implemented for real materials only at the beginning of the 2000's by Sandro Massidda and Hardy Gross' groups, a conceptually different extension of the DFT approach to the the superconducting state~\cite{Th:Luders_PRB_2005,Th:Marques_PRB_2005}.

Both AMET and SCDFT remove two strong assumptions of the simpler ME version, i.e. the
isotropy of the \ep coupling and superconducting gap and the use of an empirical 
Morel-Anderson pseudopotential.
Excellent reviews of the latest developments of the two approaches may be found in  Refs.~\onlinecite{DFT:giustino_RMP_2017}-\onlinecite{DFT:Sanna_Eliashberg_JPSJ_2018},
and
\onlinecite{DFT:Sanna_SCDFT_2017}, respectively. For our historical perspective, it is sufficient to say that the two approaches  have by now
reached comparable accuracy ($\sim 10 \%$ on the \tc), and that a formal mapping of the two theories is possible
using the Sham-Schl\"uter connection~\cite{DFT:Sham_PRL_1983}.


The first attempt to calculate the {\tc} of a
 conventional superconductor from first principles dates back to the 1970's, 
when Gaspary and Gyorffy proposed an approximate method to compute the \ep coupling in superconductors~\cite{Th:Gaspary_Gyorfyy_PRL1972}.
The deformation-potential approach by
Kahn and Allen is ten years older~\cite{Th:Kahn_Allen_PRB1984}.
Both approaches were too crude to yield
reliable results; a serious limitation
was represented by the prohibitive cost of computing phonon spectra in those years: 
they were fitted to experiments via semi-empirical
force-constant models.

A big progress came with the development of Density Functional Perturbation Theory (DFPT)~\cite{Baroni_RMP2001}. On this basis,
Savrasov and Savrasov published in 1996 the first complete set of \tc\; calculations, showing that 
DFPT, combined with a semi-phenomenological ME approach, 
was capable of a reliable description of the superconducting and transport properties
for all known elemental superconductors at ambient pressure, and also of the explanation why 
some elements, like Cu or Pd, are not superconductors~\cite{DFT:Savrasov_PRB_1996}.
The importance of this work was recognized only a few years later, because, at that time, the superconductivity
debate was dominated by the cuprates, where such a conventional \ep mechanism
 for the high-\tc\; had been clearly ruled out~\cite{SC:Savrasov_PRL_1996}.

Moreover, an old semiempirical argument based on the simplest version of the ME theory 
predicted an upper limit of 25 K for the \tc\; 
of conventional superconductors  --the so-called 
Cohen-Anderson limit-- and this seemed to be confirmed by the experimental
discoveries: year by year  \tc\; had been slowly and
painfully pushed  up to 25~K in the best superconductors known at that time, 
the A15, (Nb$_3$Ge), and it seemed impossible to go
beyond that value; cuprates were not an exception to this rule,
since their superconductivity was soon recognized as
 due to a coupling mechanism other than \ep coupling.

This is why, in our recollection of the path to room-temperature superconductivity, a much less spectacular
discovery paradoxically appears more important than that of cuprates: in 2001,
a simple intermetallic, {\em magnesium diboride} (\mg),  was found to be superconducting with 
a \tc\; of 39 K.~\cite{SC:akimitsu_mgb2}
Besides disproving the Cohen-Anderson limit for \tc, \mg\; has characteristics which
make it stand out of the group of previously known conventional superconductors: its superconducting gap has 
very different values on different sheets of its Fermi surface (two-gap superconductivity) and its
phonon spectrum is strongly anharmonic.\cite{SC:Canfield_mgb2}
Fortunately the description of such effects requires a relatively straightforward extension of the standard
theory of conventional superconductivity which can be  incorporated into ab-initio
approaches:~\cite{SC:mgb2:kong,SC:mgb2:kortus1,SC:mgb2:kortus2,SC:mgb2_choi_2002} the {\mg} case induced  important methodological developments, such
as the implementation of the anisotropic formalism for \ep interaction, the development
of Wannier-function methods to accurately compute linewidths and Kohn anomalies, 
a new formalism to compute self-consistently anharmonic effects
on phonon spectra, etc. ~\cite{DFT:Giustino_PRB_2007,DFT:profeta_PRB_2010,DFT:Errea_PRB_2014}
But the most important role played by the \mg\; discovery was probably to provide a
new playground for the understanding of material-specific
mechanisms which within the conventional \ep coupling may give rise to a high \tc.

Until 2001, the best conventional superconductors were intermetallics 
containing $d$ metals, and the main strategy adopted to ``boost'' their \tc\; was the 
use of different dopants to increase their density of states at the Fermi level. Unlike them,
{\mg} was a simple
$s$--$p$ material whose high-\tc\; derived from ``covalent bonds driven metallic'',
as nicely synthesized by the title of a paper by Warren Pickett~\cite{SC:mgb2:pickett}.

The meaning of these expressions is most easily understood if one
rewrites the \ep coupling constant $\lambda$ using the simplified
Hopfield expression:
\begin{equation}
\label{eq:lambda}
\lambda=\frac{N\hspace{-0.6 pt}(0) \hspace{0.8 pt}I^2}{M \omega^2},\,\,\text{where}
\end{equation}
$N(0)$ is the density of states at the Fermi level, $I$ the \ep matrix element, and $M \omega^2$
a phonon force constant.

In most elemental superconductors and, to a lesser extent, in A15 compounds,
the \ep coupling spreads  over several phonon modes and electronic states;
the highest \tc's are obtained from the largest $N(0)$ values
(typically within narrow transition-metal bands) combined with a moderate average {\ep} matrix element and relatively low phonon frequencies, and remain limited to $\sim$ 25 K.

On the other hand, {\mg} (and also, as we shall later see, some high-pressure hydrides) have relatively small $N(0)$'s.
What drives the high \tc, in these cases,
are large \ep matrix elements ($I$) between a
  few selected bond-stretching phonons and the electrons
  which contribute to those directional, covalent bonds.
 However,  such electrons are shared by neighboring atoms
within fully occupied bonding states which are well separated in energy from the corresponding, empty antibonding states; as a result, under normal external pressure, almost all covalent or covalent-polar solids are insulators with $N(0)\!=\!0$.
The few known exceptions to the thumb rule ``covalent materials are insulators''
all turned out to be superconductors, with higher or lower \tc\;'s primarily depending on how large or small is $N(0)$: 
 \mg\; ($\sim$40 K),
  B-doped insulators ($\sim$10 K)~\cite{Boeri_diamond_PRL2004,SC:Pickett_diamond_PRL_2004},
and also hypothetical compounds like hole-doped LiBC or graphane~\cite{SC:Rosner_LiBC_PRL2002,SC:giustino_PRL_2010}
  for which  \tc\;'s of $\sim$100~K were theoretically predicted.
Briefly, the experimental discovery of  \mg\;  
induced a fresh-mind theoretical re-examination of the balance among different
material-specific ingredients of a high-\tc\; conventional superconductor,
which in turn ignited a big hunt for
covalent and light-element superconducting compounds. In a few years, several classes of new
(conventional) superconductors were experimentally discovered and theoretically  interpreted with
{\em ab-initio} calculations~\cite{SC:diamond:Ekimov_2004,Blase_NATMAT,SC:weller_nphys_2005}. 

The above discoveries are very important, but not directly related to the recent achievement of room-temperature superconductivity, so here we will not discuss them any further.
We will, instead, shift to a topic which is at first sight
unrelated to superconductivity.

In the same years as the \mg\; discovery, major progresses took place in the field of {\em high pressure
  research}~\cite{PaulMc_NatHPMat}.
Better diamond anvil cells (DAC) equipped with innovative experimental setups 
allowed to perform in-situ resistivity, susceptibility, Raman and IR-spectra measurements
up to pressures in the Megabar range.
As a result, in the first years of the 21$^{st}$ century the phase diagram of many 
compounds was explored to an unprecedented degree of accuracy.
As more and more information on the behavior of matter 
under extreme pressure conditions piled up,
it became clearer and clearer that, under pressure, 
even the structural evolution of the simplest elemental solids
defies the chemical understanding based on ambient-pressure experience. 
Widely accepted concepts, such as the idea that, under sufficiently high pressure,
all elements eventually form close-packed structures, urgently needed to be
revised.

A 2005 survey of the superconducting properties of all elemental solids shows that, under appropriate conditions, almost all elements of the Periodic Table can be made superconducting, including those which are insulating at ambient pressure, like silicon, boron or oxygen~\cite{superconductingelements_2005}.

Sandro's group has played an important role in high-pressure superconducting research since its early days, when it appeared a mere intellectual curiosity. High-pressure experiments looked like an ideal testbed for the 
newly-developed SCDFT method.
Particularly instructive, in this respect, are his early works on lithium
and other alkali metals under pressure, which discuss in detail the
role of phonon softening and demonstrate an anomalously strong
role of the residual electronic screening in suppressing
the \tc~\cite{LI_TCP_Profeta_2006}.
Sandro's interest in high pressure remained high; his most recent work
on superconductivity in sulfur appeared in  2017~\cite{Sulfur_TCP_Monni_2017}.

A development which was crucial to interpret high-pressure
superconductivity experiments was {\em ab-initio} crystal structure
prediction (CSP), which was also developed in the same years~\cite{DFT:Woodley_Catlow_nmat_2008}.
The basic idea of CSP is to use efficient search and optimization
methods to find the global minimum of a complex energy landscape,
representing a given compound at given external conditions.
If the search space is sufficiently large, these methods often
permit to identify the actual ground-state structure of a system.
Such a  possibility to predict the crystal structure
of a material  from first principles is particularly attractive when diffraction experiments are
difficult or impossible to perform.
Two of the early successes of  {\em ab-initio} crystal structure prediction were the identification of the high-pressure superconducting phases of boron (max \tc\; 11 K)~\cite{SC:boron_science_2001} and calcium (\tc\; 25 K)
~\cite{Ca_HIGHP_EXP_2006},
two light elements with low X-Ray cross sections.
After their equilibrium crystal structures were determined by CSP methods, 
{\em ab initio} calculations reproduced their experimental \tc's 
with an accuracy of a few K,
as well as their experimental pressure behavior~\cite{Boron_highP_TC_2004,SC:Martonak_Ca_PRL2009}.

In the various ``Periodic Tables of Superconducting Elements''
published over the last ten years
the first element, {\em hydrogen}, is always missing.
The reason is practical: the pressure required to metallize hydrogen 
is at the limit of todays' capabilities.

But as soon as the (huge) metallization pressure is reached, metallic hydrogen is expected to be
a superconductor, and a good one, with very high \tc\;. The original
argument, proposed by Neil Ashcroft as early as 1968~\cite{H:Ashcroft_PRL1968}, can be understood on the basis of Eq.~\ref{eq:McMillan}: ($i$) Due to its low atomic mass, the characteristic frequencies of hydrogen are high (prefactor $\omega_{\log}$);
  ($ii$) Due to the lack of screening from core electrons, electron-ion matrix elements are high (exponent $\lambda$); this means that, even with a moderate $N(0)$, \tc's can be quite high.
Ashcroft's argument is so general, that it does not require any specific assumption on the 
crystal structure of hydrogen. Based on the simplified understanding of elemental crystals under pressure, early ab-initio calculations of the superconductivity of metallic hydrogen  typically assumed fcc structures, and
predicted \tc's as high as 600 K~\cite{H_TCfcc_Freeman_1984,H_TCfcc_maksimov_2001}.

The actual high-pressure phase diagram of hydrogen turned out to be, however, much more complex than initially thought.
A very important step in its understanding is a computational
study of the Cambridge group, published in 2007 and based on ab-initio random structure searches,
identified several high-pressure phases for which
only limited spectroscopic information was, at the time, available~\cite{H_TH_Pickard_nphys2007}.

At ambient pressure, hydrogen forms a {\em molecular} crystal in which the H$_2$ molecules are loosely arranged on a disordered lattice;  in this structure, hydrogen exhibits a gap as large as 10~eV.
Under increasing pressure, the H$_2$ molecules tend to arrange on more and more regular lattices,
giving rise to a sequence of phase transitions.
In one of these {\em molecular} structures, a band-overlap insulator-to-metal transition is predicted to occur at $\sim 400$ Gpa.
Slightly after (at $\sim 500$ GPa), another transition is predicted to occur towards an {\em atomic} ($\beta-Sn$) structure in which the molecular units are dismantled; this structure is also metallic.

In 2008, Sandro's group published what 
can be considered  the first  {\em ab-initio} calculation of the \tc\;
of metallic hydrogen in a physically-meaningful structure~\cite{H:Cudazzo_hydrogen_2008,Cudazzo_1_PRB2010,Cudazzo_2_PRB2010}.
The calculation assumes a $Cmca$ structure, a simplified
  version of the intermediate high-pressure {\em molecular} structures 
  in which one can still identify two different {\em inter} and {\em intra}-molecular bonds.
  As the length of these two types of bonds becomes comparable,
  the gap between the bonding and antibonding states closes, and
  hydrogen undergoes an insulator-to-metal transition. The resulting Fermi surface comprises both holes (bonding)
  and electron (antibonding) pockets. According to Sandro and collaborators, this should lead  
  to a complex three-band structure of the superconducting gap.
The corresponding predicted \tc's (up to 300 K at 500 GPa) were one order of magnitude higher than what had been observed in other elements, 
confirming on a quantitative basis Ashcroft's 1968 intuition that,
at least in principle, high-\tc\, conventional superconductivity was not impossible.
More recent calculations for the {\em atomic} $\beta-Sn$ phase predict an equally large
\tc's.~\cite{H:McMahon_PRB_2011,H:Borinaga_PRB_2016}

Until a few years ago the metallization of hydrogen was beyond reach, but
at least three experimental groups have reported it in the last two years, at pressures ranging from 360 to 500 GPa
(3.6 to 5 Mbar)\cite{H:Dalladay_H_Nature2016,H:Eremets_H_arxiv2016,H:Dias_science2017}; the crystal structure
of the metallic phase and hence the mechanism of metallization remain, however, controversial. 
Direct  evidence is missing and  computed transition pressures
have a large uncertainity due to  quantum lattice effects, which in hydrogen may strongly
affect the relative stability of different phases as well as the phonon spectra~\cite{H:McMahon_RMP_2012,H:errea_nature_2016}.
Aside from its experimental realization, which looks closer and closer,
the main relevance of superconducting hydrogen is not practical, 
because of the huge metallization pressure required,
but rather theoretical, as the starting idea whose developments eventually led to
the first actual discovery of conventional high-\tc\; superconductivity.
 

At the beginning of the century
{\em covalent hydrides} were intensely studied
for their ability to incorporate and release
hydrogen under appropriate conditions of temperature and
pressure~\cite{wolverton_H_CSR}.
Efficient hydrogen-storage materials are at the heart of hydrogen fuel
cells, a clean alternative to fossil fuels for on-board automotive applications.
They received a strong attention in a period when, for political
and technological reasons, it seemed highly likely that the growing
demand for fossil fuels would be hard to meet in the near future -- See
Ref. ~\onlinecite{URL_hydrogen}.

In 2004, Ashcroft conjectured that the ability of these hydrides 
to trap a large fraction of hydrogen in a host lattice could be exploited to exert
a chemical pressure on hydrogen, thus lowering its metallization
pressure~\cite{H:Ashcroft_PRL2004}.
The first experiments on silane (SiH$_4$) were perfomed in 2008
and confirmed  Ashcroft's intuition, but
\tc's were disappointingly low~\cite{H:silane_exp_2008}.
Such a low \tc\, 
did not imply
that the strategy to metallize hydrogen
using hydrides
was  wrong {\em per se}.
In hydrides, just like in other classes of superconductors,
small details of the electronic structure which
depend on crystal structure, chemical composition, doping, and other intrinsic or external conditions,
would probably have a crucial impact on  \tc\,. Superconductivity was indeed close, but the chemical composition 
yielding a high-\tc\; hydride was not as trivial as originally hoped.

At this point, the experience gained with elemental solids at high pressure turned out to
be extremely precious: it had been convincingly demonstrated that, in fact,  {\em ab-initio}
methods for crystal structure prediction, combined with methods to compute the critical
temperatures of superconductors, could be used to calculate accurate phase diagrams.
Hydrides, being binary systems, pose the additional complication that several compositions
may form and cohexist as a function of pressure;
however, estimating the relative stability of 
different compositions by first-principles calculations through the {\em convex hull} constructution is rather straightforward~\cite{ceder_phase_diag}.

In the years immediately following Ashcroft's prediction, many binary hydrides were computationally investigated, in hope to identify prospective
high-\tc\; superconductors.
A very important milestone  in this process is a paper by Zurek et al., published in 2009, where, for the first time, it was explicitly
pointed out that off-stoichiometry phases of hydrides,  in particular the {\em superhydrides} which form at
high pressure, exhibit very different properties from their ambient-pressure counterparts, and some of them also show high-\tc\, superconductivity~\cite{H:zurek_PNAS_2009}.

The real breakthrough came in 2015, when, for the first time, a \tc\; exceeding 200 K was measured in pressurized \sh, which
 forms  when sulfur di-hydride (SH$_2$) is placed in a hydrogen-rich
atmosphere and compressed in a DAC above $\sim 10$ GPa~\cite{DrozdovEremets_Nature2015}.
The H$_2$S molecule, which is analogous to water,  at ambient pressure forms a molecular, insulating crystal; in the low-pressure phases of \sh,
 also  insulating,
hydrogen is trapped in molecular form  (H$_2$) in the open H$_2$S lattice.
As pressure is increased above 20 GPa, hydrogen
is gradually incorporated into the SH$_2$ lattice, and, at $\sim$ 100 GPa,
\sh\; units start to form. This causes a gradual insulator-to-metal transition.
At 150 GPa the S-H interaction becomes so strong that \sh\; orders in a highly-symmetric ($bcc$) structure with a rather unusual bonding: sulfur forms covalent bonds with its three hydrogen neighbours.
In analogy to \mg\,, these bonds, driven metallic, are at the origin of the high-\tc\; conventional superconductivity of \sh~\cite{H:Bernstein_PRB_2015,mine:Heil_PRB_2015}.

The discovery of high-\tc\; superconductivity in \sh\; is an impressive demonstration of the progress of current high-pressure experimental techniques, but it also
represents the 
first unquestionable breakthorugh of the {\em ab-initio} approach to the search
and discovery of
new superconductors: it was, in fact,
a theoretical DFT paper~\cite{H:Duan_SciRep2014}
which, a few
months before the experimental report, 
predicted the exact pressure, structure, \tc\; and chemical
composition of superconducting \sh.

This was not the only succesful prediction of high-\tc\; superconductivity in high-pressure hydrides. A few months later, several theoretical
papers pointed out that the neighbour of sulfur in the Periodic Table,
phosphorus, should also form superconducting hydrides at high pressures, with \tc's of $\sim 100$ K, 
but also that all these superconducting phases would be metastable with respect to elemental decomposition~\cite{H:Fu_Ma_pnictogenH_2016,PH3_Eremetsprivate, Flores_PH3_PRBR2016}.
The metastability of such samples was later confirmed experimentally~\cite{H:Drozdov_PH3_arxiv2015,PH3_Eremetsprivate}.

Even more impressive is the case of lanthanum: in 2017 a computational study 
predicted that it forms superconducting {\em superhydrides} with \tc's
as large as 300$~K$~\cite{MA_REHX_PRL_2017}, and in the summer of 2018 two different groups reported that
one of these superhydrides, \lah, does indeed form in a DAC, and
their highest reported \tc\, (265~K) is close to room temperature.

\lah\; 
belongs to a larger class of sodalite-like clathra\-tes
of chemical formula $X$H$_n$
which, according to theoretical predictions, should also include the hydrides of
yttrium, scandium, magnesium, calcium and
several rare earths~\cite{H:Wang_PNAS2012_CaH6,MA_REHX_PRL_2017}.
In these compounds hydrogen forms interconnected polyhedral cages
each enclosing a guest atom $X$, yielding 
a dense sponge-like hydrogen lattice in which the shortest H-H distances 
are close to those predicted for metallic atomic hydrogen in the $\beta$-Sn structure. In fcc  \lah\,, for example, each lanthanum sits in the middle of a 32-hydrogen cage; in bcc LaH$_6$, the lanthanum's nearest neighbors are 24 hydrogen atoms; and so on.
Many such  {\em superhydrides}
with a sodalite-like clathrate structure 
have been predicted to exhibit high-\tc\, conventional superconductivity,
the highest \tc's being expected for yttrium, calcium and magnesium~\cite{zurek_review}.

After the spectacular demonstration of their predictive power 
not only for the new high-\tc\, compound \sh\,, but also for
the possibility of room-temperature conventional superconductivity
in \lah\,, a natural question to ask is: what is the next goal of electronic-structure methods?

In our view, the first aim is to understand {\em why} some hydrides achieve
room-temperature superconductivity, while others don't.
This is a step beyond the state-of-the-art ability of computing 
and successfully predicting phase diagrams and superconductive
\tc's of different hydrides from first principles, and
amounts to identifying the specific mechanisms leading
to high-\tc\; in different hydrogen compounds.
 
First of all, one has to acknowledge
the evidence of qualitatively different families
within the high-pressure hydrides.
For example, the high-\tc\, superconductivity mechanism cannot be the same
in \sh\, and \lah\,: at variance with \sh\,, where sulfur is incorporated into the H sublattice, forming covalent S-H bonds, the main role of the guest atom in sodalite-like clathrates is to provide charge  and stabilize the 
the H sublattice by ``filling the holes'' in the sponge-like geometry.

Then one has to recognize that H-H distances close
to those realized in atomic $\beta$-Sn hydrogen
are a necessary, but not sufficient condition for
high-\tc\, superconductivity.
For example, when first synthesized,
the crystal structure
of FeH$_5$ (which can  be seen as a cubic FeH$_3$ lattice
intercalated by two-dimensional H layers),
seemed a promising candidate for high-\tc\; superconductivity~\cite{H:Pepin_FEH_2017},
 because of the propitious H-H distances in the intermediate layer;
first-principles calculations
demonstrated that, instead, the Fe-H bonds which dominate the electronic states at the Fermi level
are too weak to support it~\cite{H:Heil_FEH_2018}.

The second aim, directly connected to the first, is
to understand whether and how superconductivity in high-pressure hydrides can
have any impact on actual applications.
This obviously means to seek and achieve
a substantial {\em reduction of the metallization and superconducting pressure},
a goal which may be sought according to different strategies:
\begin{itemize}
\item
already known hydrides may be used a starting point for optimization strategies.
For example, in sodalite-like clathrates, the evident correlation
between the size and valence
of the guest atom and the stabilization pressure
may be systematically studied and
exploited to lower the effective metallization pressure~\cite{mine:Heil_CM2019};
\vspace{-0.2 cm}
\item
ternary hydrides may offer an even larger flexibility to stabilize 
different crystalline superhydride phases~\cite{mine:Kokail_PRM_2017},
or to achieve metallization at lower pressure via doping of 
molecular phases~\cite{H:Flores_H2O,Flores_CH_EPJB_2018};
\vspace{-0.5 cm}
\item
compounds containing light elements other than hydrogen,
		although their high-pressure behavior can be quite different 
		from hydrides,~\cite{naumov2015chemical}
		may also be made superconducting under high pressure~\cite{Kokail_Li3S_PRB2016}; among them the elements forming covalent bonds at ambient pressure,
like boron and carbon, are particularly promising, since, when doped,
these bonds have already proved to support high \tc. 
As a matter of fact, the current record for conventional superconductivity at 
ambient pressure (56$~K$) has been achieved
last year in amorphous boron-doped carbon~\cite{SC:qcarbon_2017},
and even higher \tc's have been predicted in (ambient-pressure) doped hydrogenated graphene (graphane)~\cite{SC:giustino_PRL_2010}.
\end{itemize}
In summary, first-principles calculations
of the superconducting properties of conventional superconductors, 
a field of research pioneered by Sandro Massidda,
have by now gained the status of a reliable and 
efficient tool not only to interpret existing superconductors, but 
also to predict new ones, 
and are likely to play a decisive role in the strategies
aimed at achieving ambient-pressure high-\tc\, superconductors.

\begin{acknowledgments}
 The authors acknowledge support from Fondo Ateneo-Sapienza 2017.
\end{acknowledgments}

\bibliographystyle{apsrev4-1}
%

\end{document}